\documentclass[aps,pra,preprint,footinbib,superscriptaddress]{revtex4-1}
\usepackage{graphicx}
\usepackage{indentfirst}
\usepackage{braket}
\usepackage{float}

\usepackage{epsfig,amssymb, amsmath, mathtools}
\newcommand{\bp}{\boldsymbol \rho}
\newcommand{\bpt}{\boldsymbol{\tilde{\rho}}}
\newcommand{\bpp}{\boldsymbol{\rho}_+}
\newcommand{\bpm}{\boldsymbol{\rho}_-}
\newcommand{\bmu}{\boldsymbol \mu}

\newcommand{\bt}{\boldsymbol{\theta}}
\newcommand{\bk}{{\bf k}}
\newcommand{\bs}{{\bf s}}
\newcommand{\bsp}{\bs_+}
\newcommand{\bsm}{\bs_-}
\newcommand{\bst}{\boldsymbol{\tilde{\mathrm{s}}}}

\newcommand{\scP}{{\mathcal P}}

\newcommand{\scE}{{\mathcal E}}
\newcommand{\scEt}{\tilde{\mathcal E}}

\newcommand{\scS}{{\mathcal S}}

\newcommand{\scI}{{\mathcal I}}
\newcommand{\scR}{{\mathcal R}}

\newcommand{\wo}{\omega_0}
\newcommand{\ko}{k_0}
\newcommand{\kot}{\tilde{k}_0}
\newcommand{\lo}{\lambda_0}
\newcommand{\wm}{\omega_-}
\newcommand{\wpp}{\omega_+}

\newcommand{\dw}{{\Delta\omega}}
\newcommand{\nm}{\,\text{nm}}

\newcommand{\intI}[1]{\int\!{\rm d}#1\,}
\newcommand{\intII}[1]{\int\!{\rm d}^2#1\,}

\newcommand{\intw}[1]{\int\!\frac{{\rm d}#1}{2\pi}\,}
\newcommand{\intk}[1]{\int\!\frac{{\rm d^2}#1}{(2\pi)^2}\,}
\newcommand{\avg}[1]{\langle #1 \rangle}

\newcommand{\abs}[1]{\left| #1 \right|}
\newcommand{\bracket}[1]{\left[ #1 \right]}
\newcommand{\parens}[1]{\left( #1 \right)}

\begin{document}
\title{Nonparaxial phasor-field propagation}

\author{Justin Dove}
\email{dove@mit.edu}
\author{Jeffrey H. Shapiro}
\email{jhs@mit.edu}
\address{Research Laboratory of Electronics, Massachusetts Institute of Technology, Cambridge, MA 02139, USA}

\begin{abstract}
	Growing interest in non-line-of-sight (NLoS) imaging, colloquially referred to as ``seeing around corners'', has led to the development of phasor-field ($\scP$-field) imaging, wherein the field envelope of amplitude-modulated spatially-incoherent light is manipulated like an optical wave to directly probe a space that is otherwise shielded from view by  diffuse scattering. Recently, we have established a paraxial theory for $\scP$-field imaging in a transmissive geometry that is a proxy for three-bounce NLoS imaging [J. Dove and J. H. Shapiro, Opt. Express {\bf 27}(13) 18016--18037 (2019)].  Our theory, which relies on the Fresnel diffraction integral, introduces the two-frequency spatial Wigner distribution (TFSWD) to efficiently account for specularities and occlusions that may be present in the hidden space and cannot be characterized with $\scP$-field formalism alone.  However, because the paraxial assumption is likely violated in many, if not most, NLoS scenarios, in the present paper we overcome that limitation by deriving a nonparaxial propagation formula for the $\scP$ field using the Rayleigh--Sommerfeld diffraction integral. We also propose a Rayleigh--Sommerfeld propagation formula for the TFSWD and provide a derivation that is valid under specific partial-coherence conditions. Finally, we report a pair of differential equations that characterize free-space TFSWD propagation without restriction.
\end{abstract}

\maketitle

\section{Introduction}

Over the last decade, considerable progress has been made on the daunting task of seeing around corners by collecting and processing light reflected from a hidden scene into a visible space. This field of research, as well as the analogous one of seeing through diffusive transmission media like ground glass or fog, is called non-line-of-sight (NLoS) imaging~\cite{Kirmani2011,Velten2012,O'Toole2018,lindell,Xu2018,Thrampoulidis2018}. One of the more promising developments in this line of work is phasor-field ($\scP$-field) imaging. Initially proposed by Reza \emph{et al.}~\cite{Reza2018} and demonstrated by Liu \emph{et al.}~\cite{Liu2019,Liu2020}, the key idea behind $\scP$-field imaging is to amplitude modulate a visible light source at radio or microwave frequencies and to manipulate, detect, and process the modulation envelope, i.e., the short-time-average (STA) irradiance, as if it were an optical wave using techniques from traditional line-of-sight (LoS) imaging. The difficulty in NLoS imaging arises from ordinary walls and diffusive transmission media being rough at the optical-wavelength scale, resulting in diffusely-scattered, spatially-incoherent light. However, at the much larger scale of radio and microwave wavelengths, i.e., the modulation wavelength of the $\scP$ field, these walls and transmission media are smooth, and so the $\scP$ field retains modulation-frequency directionality information in a manner similar to that for coherent light at the optical frequency in LoS scenarios.  

Although the $\scP$-field concept proved intuitively pleasing and experimentally impressive, we felt there was room for a sounder understanding of its theoretical underpinnings. So, contemporaneously with Teichman~\cite{Teichman2019}, we began a program of research dedicated to pursuing such understanding. Thus far the fruit of that labor is a theoretical framework for $\scP$-field propagation through transmissive geometries which serve as proxies for more typical reflective NLoS scenarios~\cite{Dove2019}. Our framework is capable of characterizing both computational~\cite{Dove2019} and physical-optics $\scP$-field imaging~\cite{lenses}, efficiently accounting for hidden-space specularities and occlusions by means of the two-frequency spatial Wigner distribution (TFSWD)~\cite{Dove2019}, as well as analyzing the effects of laser speckle~\cite{speckle}. Our framework, however, is limited to the paraxial, i.e., small-angle, propagation regime as it heavily relies on the Fresnel diffraction integral, whereas other formulations started from the more general Rayleigh--Sommerfeld diffraction integral \cite{Reza2018,Teichman2019}. 

The paraxial approximation is valid for propagation of an optical wave at wavelength $\lo$ across a distance $L$ when the transverse diameters of interest at the input and output planes, $d_0$ and $d_L$ respectively, satisfy $(d_0+d_L)^4/128\lo L^3 \ll 1$. However, typical parameter values for many NLoS imaging scenarios violate this condition, sometimes greatly~\cite{Liu2019}. An alternative sufficient condition for the validity of Fresnel diffraction is that the input field's angular spectrum be comprised solely of components that lie close to the axis of propagation. Unfortunately, insofar as this alternative is concerned, the defining characteristic of NLoS scenes is their involving rough surfaces that scatter light over a broad range of angles. Accordingly, it seems that the paraxial approximation is likely to be violated in many, if not most, NLoS scenarios.  Thus our paraxial theory of $\scP$-field propagation should be extended to a nonparaxial theory by replacing our assumption of Fresnel diffraction with the more broadly valid Rayleigh--Sommerfeld diffraction.  This is especially true as we hope to adapt our transmissive framework to the reflective geometry of real NLoS problems. In this paper, we take initial, key steps towards that goal.

First, in Sec.~\ref{sec:setup}, we review our paraxial $\scP$-field propagation framework both for its own merit and to remotivate our introduction of the TFSWD to handle hidden-space specularities and occlusions. Then, in Sec.~\ref{sec:pfield}, we use the Rayleigh--Sommerfeld diffraction integral to derive the appropriate $\scP$-field propagation integral for free-space propagation following a diffuser and show the equivalent implication for the propagation of the diffuser-averaged STA irradiance. In Sec.~\ref{sec:tfswd} we call attention to the paraxial free-space propagation primitive for the TFSWD and show its equivalent formulation in terms of the 6D light field, viz., the TFSWD's temporal Fourier transform. Using geometric intuition, we propose a nonparaxial propagation primitive for the 6D light field by replacing the paraxial terms in our previous result with their nonparaxial equivalents. We verify this proposal implies the correct behavior for the diffuser-averaged STA irradiance, and exhibit its equivalent formulation in terms of the TFSWD.  We then provide a more formal derivation of that TFSWD equivalent which is valid under specific partial-coherence conditions. Finally, in Sec.~\ref{sec:helmholtz}, we provide a pair of differential equations that characterize free-space TFSWD propagation without restriction.

\section{Paraxial $\scP$-field and TFSWD propagation}
\label{sec:setup}
\begin{figure}[hbt]
	\centering
	\includegraphics[width=4.5in]{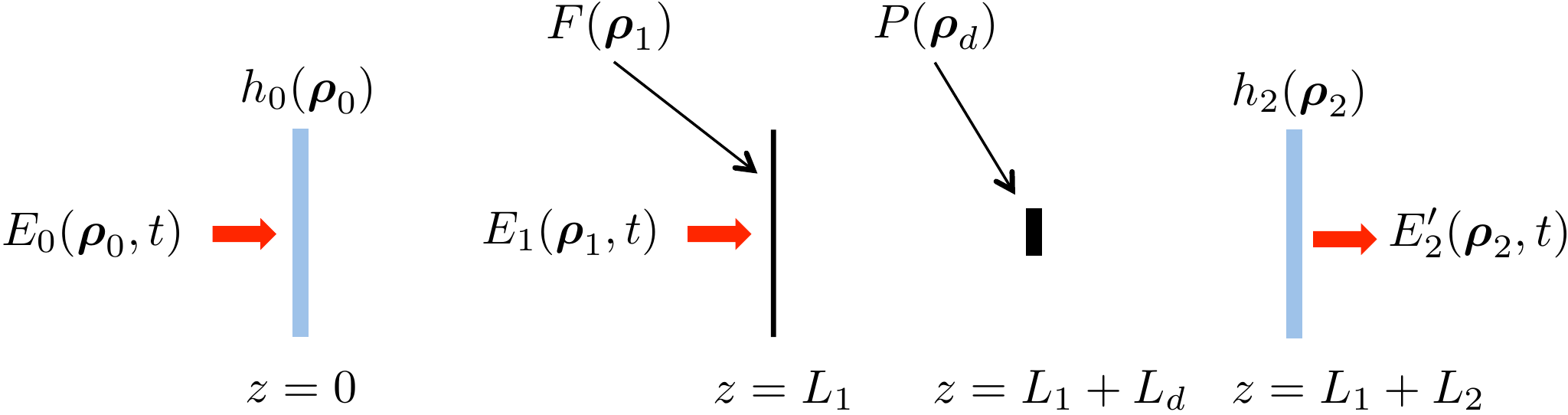}
	\caption{Transmissive geometry that serves as an unfolded proxy for a three-bounce NLoS scenario whose hidden space contains a specular component and an occluder. The thin blue rectangles represent idealized thin diffusers that serve as analogs for the first and final bounces off the visible wall. The black line in the central plane represents a target in the hidden space with both specular and diffuse components. The small black rectangle between this target and the final plane represents an occluder in the hidden space that partially obstructs the light returning to the visible wall.\label{fig:geo}}
\end{figure}

To begin, we review the paraxial phasor-field propagation framework that we developed in Ref.~\cite{Dove2019}. We use paraxial, scalar-wave optics in a transmissive geometry that serves as an unfolded proxy for three-bounce NLoS imaging, as depicted in Fig.~\ref{fig:geo}. Our framework characterizes propagation through such geometries in a plane-by-plane fashion, i.e., with a set of primitives that describe interactions with the variety of planar elements displayed in Fig.~\ref{fig:geo} and propagation from one non-empty plane to the next. Our paraxial assumption only enters into the propagation primitives, so it is only these primitives that need modification to extend our framework to the nonparaxial regime. 

The fundamental quantity we concern ourselves with is the optical field's baseband, ${\rm W}^{1/2}/{\rm m}$-units, complex field envelope, $E_z(\bp_z,t)$, which modulates a $\wo$-angular-frequency optical carrier to produce the optical field, $U_z(\bp_z,t)=\text{Re}[E_z(\bp_z,t)e^{-i\wo t}]$, where $\bp_z$ is the two-dimensional transverse spatial coordinate in the plane denoted by $z$~\cite{footnote-z}. We take the bandwidth of this envelope to satisfy $\dw \ll \wo$ and its frequency-domain representation to be $\scE_z(\bp_z,\omega) = \intI{t} E_z(\bp_z,t) e^{i\omega t}$. In the frequency domain, propagation of this envelope through our transmissive geometry is characterized by the following primitives:

\noindent{\bf Propagation through diffusers: }The frequency-domain field envelopes $\{\scE'_k(\bp_k,\omega) : k = 0,2\}$ emerging from the $z=0$ and $z=L_1+L_2$ plane diffusers are given by
\begin{align}
	\scE_k'(\bp_k,\omega) = \scE_k(\bp_k,\omega) e^{i(\wo+\omega) h_k(\bp_k)/c} \approx \scE_k(\bp_k,\omega)e^{i\wo h_k(\bp_k)/c}, 
\end{align}
where $c$ is light speed and the $\{h_k(\bp_k)\}$, which represent the diffusers' thickness profiles, are  independent, identically-distributed, zero-mean Gaussian random processes with standard deviation satisfying $ \lo \ll \sigma_h \ll \Delta\lambda$ and correlation width obeying $\rho_h \sim \lo$, with $\lo \equiv 2\pi c / \wo$ being the optical wavelength and $\Delta\lambda\equiv 2\pi c / \dw$ the minimum modulation wavelength. It follows that the diffuser correlation function appearing in propagation integrals, such as Eq.~(\ref{propeqn}) below, reduces to 
\begin{align}
	\avg{e^{i\wo[h_k(\bp_k)-h_k(\bp_k')]/c}} = e^{-\wo^2[\sigma_h^2-K_{hh}(|\bp_k-\bp_k'|)]/c^2} \approx \lo^2 \delta(\bp_k-\bp_k'),\label{eq:diff-stat}
\end{align}
where angle brackets denote ensemble average, $K_{hh}(|\bp_k-\bp_k'|) \equiv \langle h_k(\bp_k)h_k(\bp_k')\rangle$ has been assumed to be homogeneous and isotropic, and $\delta(\cdot)$ is the Dirac delta function.

\noindent{\bf Propagation through transmissivity masks: }The frequency-domain field envelope $\scE'_d(\bp_d,\omega)$ emerging from the deterministic occluder is given by
\begin{align}
	\scE_d'(\bp_d,\omega) = \scE_d(\bp_d,\omega) P(\bp_d),
\end{align}
where $P(\bp_d)$ is the occluder's field-transmissivity function.  Similarly, the frequency-domain field envelope $\scE'_1(\bp_1,\omega)$ emerging from the target plane is given by 
\begin{align}
	\scE_1'(\bp_1,\omega) = \scE_1(\bp_1,\omega) F(\bp_1).
\end{align}
Here, $F(\bp_1)$ is a random process having potentially-nonzero mean $\langle F(\bp_1)\rangle \neq 0$, representing a specular component.  The mask's fluctuations, $\Delta F(\bp_1) \equiv F(\bp_1)-\langle F(\bp_1)\rangle$, represent its diffuse component and have covariance function ${\langle \Delta F(\bp_+ + \bp_-/2)\Delta F^*(\bp_+ - \bp_-/2)\rangle} \approx \lambda_0^2\mathcal{F}(\bp_+)\delta(\bp_-)$, where $0\le \mathcal{F}(\bp_+) \le 1$.

\noindent{\bf Fresnel diffraction: }Fresnel diffraction governs free-space propagation from $z=0$ to $z=L_1$, viz., 
\begin{align}
	\scE_1(\bp_1,\omega) = \frac{\wo+\omega}{i2\pi c L_1} \intII{\bp_0} \scE_0'(\bp_0,\omega)\exp\!\bracket{i(\wo+\omega)\parens{L_1+|\bp_1-\bp_0|^2/2L_1}/c}.
	\label{eq:E-fresnel}
\end{align}
Similar results relate $\scE_d(\bp_d,\omega)$ to $\scE_1'(\bp_1,\omega)$, etc. We will often use $\wo+\omega\approx\wo$ in these expressions' leading terms.

The STA irradiance is the squared magnitude of the complex field envelope, $I_z(\bp_z,t)=|E_z(\bp_z,t)|^2$, and we define the $\scP$ field to be its diffuser-averaged Fourier transform:
\begin{align}
	\scP_z(\bp_z,\wm) &\equiv \intI{t} \avg{I_z(\bp_z,t)}e^{i\wm t}\\
	&= \intw{\wpp} \avg{\scE_z(\bp_z,\wpp+\wm/2)\scE^*_z(\bp_z,\wpp-\wm/2)},
	\label{PfieldFromEfield}
\end{align}
where Eq.~(\ref{PfieldFromEfield}) follows from the STA irradiance's definition and the convolution-multiplication theorem.
In Ref.~\cite{Dove2019} we found that post-diffuser Fresnel propagation of the $\scP$ field is characterized by
\begin{align}
	\scP_1(\bp_1,\wm) = \frac{1}{L_1^2}\intII{\bp_0}\scP_0(\bp_0,\wm)\exp\!\bracket{i\wm\parens{L_1+|\bp_1-\bp_0|^2/2L_1}/c},
\end{align}
whose mimicking the complex field envelope's Fresnel diffraction integral, Eq.~(\ref{eq:E-fresnel}), is the bedrock of $\scP$-field optics.

Using the $\scP$ field alone, propagation past the target plane in Fig.~\ref{fig:geo} can only be accomplished if the target is purely diffuse, i.e., if $\avg{F(\bp_1)}=0$, in which case we have
\begin{align}
	\scP_d(\bp_d,\wm) = \frac{1}{L_d^2}\intII{\bp_1}\scP_1(\bp_1,\wm) \mathcal{F}(\bp_1)\exp\!\bracket{i\wm\parens{L_d+|\bp_d-\bp_1|^2/2L_d}/c}.
\end{align}
This is so because the spatial coherence that accrues in diffraction from $z=0$ to $z=L_1$ is \emph{not} accounted for in $\scP_1(\bp_1,\omega_-)$ but is at least partially retained in $\scE'_1(\bp_1,\omega)$ and hence must be accounted for in propagation from $z=L_1$ to $z=L_1+L_d$.   The same situation arises with the occluder, i.e., despite $\scP_d'(\bp_d,\wm) = \scP_d(\bp_d,\wm) |P(\bp_d)|^2$ being the $\scP$-field associated with $\scE_d'(\bp_d,\omega)$, propagation from $z=L_1+L_d$ to $z=L_1+L_2$ cannot be characterized with the $\scP$ field alone. Nevertheless, even with specularities or occlusions present in the hidden space, it is always possible to write down a cumulative $\scP$-field input-output relation, i.e., between $\scP_0(\bp_0,\wm)$ and $\scP_2'(\bp_2,\wm)$ in Fig.~\ref{fig:geo}, for scenarios that contain a pure-diffuser at their input.  That relation, however, cannot be found from plane-by-plane $\scP$-field propagation. For that plane-by-plane propagation the TFSWD provides an efficient solution.  

All of the elements encountered by light in a typical NLoS scenario will effect linear transformations upon the complex field envelope. Moreover, phasor-field imaging is primarily concerned with second moments of the complex field envelope~\cite{footnote-speckle}. Consequently, for plane-by-plane propagation through NLoS scenarios for the sake of recovering $\scP$-field or diffuser-averaged-STA-irradiance input-output behavior it suffices to follow the propagation of the complex field envelope's space-time autocorrelation function,
\begin{align}
	\Gamma_z(\bp_z,\bpt_z,t,\tilde{t}) \equiv \avg{E_z(\bp_z,t)E_z^*(\bpt_z,\tilde{t})}.
\end{align}
Bearing this in mind, in Ref.~\cite{Dove2019} we introduced the TFSWD,
\begin{align}
	W_{\mathcal{E}_z}(\bp_+,\bk,\omega_+,\omega_-) \equiv \int\!{\rm d}^2\bp_-\,\langle \mathcal{E}_z(\bp_+ + \bp_-/2,\omega_+ + \omega_-/2)\mathcal{E}_z^*(\bp_+ - \bp_-/2,\omega_+ - \omega_-/2)\rangle e^{-i\bk\cdot \bp_-},
	\label{TFSWDdefn}
\end{align}
which has an invertible, Fourier-transform relationship with the space-time autocorrelation and from which the $\scP$ field can be readily obtained via
\begin{align}
\mathcal{P}_z(\bp_+,\omega_-) = \int\!\frac{{\rm d}\omega_+}{2\pi}\int\!\frac{{\rm d}^2\bk}{(2\pi)^2}\,W_{\mathcal{E}_z}(\bp_+,\bk,\omega_+,\omega_-).
	\label{eq:p-from-tfswd}
\end{align}
In these equations, $\bk$ represents the transverse component of the wave vector, which characterizes  directionality information that the TFSWD possesses but the $\scP$ field lacks. This directionality information enables the TFSWD to handle the coherence accrued from free-space propagation and thus suffice to characterize plane-to-plane propagation in scenarios such as Fig.~\ref{fig:geo} that contain specularities or occlusions. Analogously, the exposure of the additional, center frequency coordinate, $\omega_+$, enables the TFSWD to characterize propagation through linear time-invariant (LTI) filters with memory or linear time-varying filters~\cite{footnote-lenses}. Owing to their invertible relationship, the TFSWD contains the same information as the space-time autocorrelation. By contrast, the $\scP$ field does not---the space-time autocorrelation and TFSWD each suffice to calculate the $\scP$ field, but not vice versa. We chose to work with the TFSWD over the space-time autocorrelation because of its similarity to the standard optical Wigner function, which has been well studied and found to be widely useful~\cite{Alonso2011,footnotenew}, and because of its simple relationship to the intuitive, radiometric quantities used in computer vision, e.g., the light field or plenoptic function~\cite{lf1,lf2,lf3}. A planar 6D light field, characterizing the amount of light of a specified optical frequency passing through a given plane at a specified point in a specified direction at a specified time, can be defined by inverse Fourier transforming the TFSWD's final coordinate:
\begin{align}
	I_z(\bpp,\bs,\wpp,t) &\equiv \frac{1}{\lo^2}\intw{\wm} W_{\scE_z}(\bpp,2\pi\bs/\lo,\wpp,\wm) e^{-i\wm t},
\end{align}
where $\bs=\lo\bk/2\pi=\bk/k_0$ is the transverse component of the unit vector pointing in the nominal propagation direction. Radiometric quantities of lesser dimension can be obtained by integrating out the undesired coordinates or transforming them as necessary, e.g., as in Eq.~(\ref{eq:p-from-tfswd}). See Appendix~\ref{appendix} for an example of the 6D light field and its relation to other quantities of interest in this paper.  

In effect, the TFSWD and 6D light field serve as augmented versions of the $\scP$ field and diffuser-averaged STA irradiance, respectively. Both contain equivalent information to the space-time autocorrelation, and we will work with each as convenient. Thus, they provide a means for developing the sort of modular primitives we seek, for any NLoS scenario of interest, while making the process of recovering a cumulative $\scP$-field input-output from such primitives as painless as possible and, in particular, much easier than doing so from the autocorrelation function or, worse, the complex field envelope directly. That is, in our opinion, they are the simplest and most intuitive quantities, in terms of their relation to the $\scP$ field, that suffice to handle the broadest class of NLoS scenarios.

In Ref.~\cite{Dove2019} we derived the following propagation primitives for the TFSWD:

\noindent{\bf Propagation through diffusers: }
\begin{align}
W_{\mathcal{E}'_k}(\bp_+,\bk,\omega_+,\omega_-) = \lambda_0^2\int\!\frac{{\rm d}^2\bk'}{(2\pi)^2}\,W_{\mathcal{E}_k}(\bp_+,\bk',\omega_+,\omega_-),\label{diffuserprim}
\end{align}
for $k \in \{0,2\}$.

\noindent{\bf Propagation through deterministic occluders: }
\begin{align}
W_{\mathcal{E}'_{d}}(\bp_+,\bk,\omega_+,\omega_-) = 
\int\!\frac{{\rm d}^2\bk'}{(2\pi)^2}\,W_{\mathcal{E}_{d}}(\bp_+,\bk',\omega_+,\omega_-)W_P(\bp_+,\bk-\bk'),
\label{occluderprim}
\end{align}
where $W_P(\bp_+,\bk) \equiv $ $\int\!{\rm d}^2\bp_-\,P(\bp_+ + \bp_-/2)P^*(\bp_+ - \bp_-/2)e^{-i\bk\cdot\bp_-}$ is the traditional Wigner distribution for the occluder's field-transmissivity pattern.

\noindent{\bf Propagation through specular-plus-diffuser masks: }
\begin{align}
W_{\mathcal{E}'_{1}}(\bp_+,\bk,\omega_+,\omega_-) =& 	\int\!\frac{{\rm d}^2\bk'}{(2\pi)^2}\, W_{\mathcal{E}_{1}}(\bp_+,\bk',\omega_+,\omega_-)W_{\langle F\rangle}(\bp_+,\bk-\bk') \nonumber\\ &+
\lambda_0^2\mathcal{F}(\bp_+)\int\!\frac{{\rm d}^2\bk'}{(2\pi)^2}\,W_{\scE_{1}}(\bp_+,\bk',\omega_+,\omega_-).
\label{maskprim}
\end{align} 

\noindent{\bf Fresnel diffraction: }
\begin{align}
	W_{\scE_1}(\bp_+,\bk,\wpp,\wm) = W_{\scE'_{0}}(\bp_+-L_1\bk/k_0,\bk,\wpp,\wm) e^{i\wm L_1\parens{1+\abs{\bk}^2/2k_0^2}/c},
\label{Fresnelprim}
\end{align}
and similarly for all other free-space propagation paths in Fig.~\ref{fig:geo}.

Although not shown in Ref.~\cite{Dove2019}, and not exploited thus far in our work, it is not hard to extend that paper's derivation of Eq.~(\ref{occluderprim}) to show

\noindent{\bf Propagation through deterministic, planar LTI filters: }
\begin{align}
W_{\mathcal{E}_H}(\bp_+,\bk,\omega_+,\omega_-) = 
\int\!\frac{{\rm d}^2\bk'}{(2\pi)^2}\,W_{\mathcal{E}}(\bp_+,\bk',\omega_+,\omega_-)W_H(\bp_+,\bk-\bk',\omega_+,\omega_-),
\label{LTIprim}
\end{align}
where $\scE_H(\bp,\omega)=\scE(\bp,\omega)H(\bp,\omega)$ is the complex field envelope emerging from a deterministic, planar LTI filter with frequency response $H(\bp,\omega)$ when illuminated by the complex field envelope $\scE(\bp,\omega)$.

Most of these propagation primitives, namely those for propagation of the complex field envelope and TFSWD through diffusers and transmissivity masks, apply in general, without any assumption about the propagation angles, as they describe transformations at a single transverse plane. So, these elements already apply to nonparaxial operation, and accordingly we need only extend our Fresnel-diffraction results for the $\scP$ field and TFSWD to the nonparaxial regime to have a complete set of nonparaxial primitives for Fig.~\ref{fig:geo}-like scenarios.

\section{Rayleigh--Sommerfeld $\scP$-field propagation}
\label{sec:pfield}
Our development of a nonparaxial propagation primitive for the $\scP$ field parallels that for Fresnel propagation as in Ref.~\cite{Dove2019}. We consider propagation first through a diffuser and then through a free-space distance $L_1$, as depicted in the first portion of Fig.~\ref{fig:geo}. The key difference in the derivation is to replace Eq.~(3) of that paper (Eq.~(\ref{eq:E-fresnel}) here)---the Fresnel diffraction integral for the complex field envelope---with the Rayleigh--Sommerfeld diffraction integral~\cite{Goodman}
\begin{equation}
	\scE_1(\bp_1,\omega) = \intII{\bp_0}\scE'_0(\bp_0,\omega)\frac{\exp\!\left[i(\omega_0+\omega)\sqrt{L_1^2 + |\bp_1-\bp_0|^2}/c\right](\omega_0+\omega)L_1}{i2\pi c(L_1^2 + |\bp_1-\bp_0|^2)},
\label{eq:RS1}
\end{equation}
where we have substituted $L_1/\sqrt{L_1^2 + |\bp_1-\bp_0|^2}$ in place of the cosine obliquity factor.  The derivation then proceeds as in the Fresnel case, utilizing the same assumptions regarding the diffuser statistics, viz., Eq.~(\ref{eq:diff-stat}), and the modulation bandwidth:
\begin{align}
	\scP_1(\bp_1,\omega_-) =& \intw{\wpp} \avg{\scE_1(\bp_1,\wpp+\wm/2)\scE^*_1(\bp_1,\wpp-\wm/2)} \\
	=& \parens{\frac{\wo}{2\pi c}}^2 \intw{\wpp} \intII{\bp_0} \intII{\bpt_0}\scE_0(\bp_0,\wpp+\wm/2)\scE^*_0(\bpt_0,\wpp-\wm/2) \nonumber\\&\times\avg{e^{i\wo[h_0(\bp_0)-h_0(\bpt_0)]/c}}  \frac{ e^{i \parens{\parens{\wo+\wpp+\wm/2}\sqrt{L_1^2+\abs{\bp_1-\bp_0}^2}-\parens{\wo+\wpp-\wm/2}\sqrt{L_1^2+\abs{\bp_1-\bpt_0}^2}}/c}L_1^2}{(L_1^2+\abs{\bp_1-\bp_0}^2)(L_1^2+\abs{\bp_1-\bpt_0}^2)} \label{propeqn}\\
	=& \intII{\bp_0} \scP_0(\bp_0,\omega_-)\frac{\exp\!\left(i\omega_-\sqrt{L_1^2 + |\bp_1-\bp_0|^2}/c\right)L_1^2}{(L_1^2 + |\bp_1-\bp_0|^2)^2}.
\label{P1eqn}
\end{align}

Recalling that the $\scP$ field is the Fourier transform of the diffuser-averaged STA irradiance, it immediately follows that 
\begin{equation}
	\langle I_1(\bp_1,t)\rangle = \intII{\bp_0} I_0\!\left(\bp_0,t-\sqrt{L_1^2 + |\bp_1-\bp_0|^2}/c\right)L_1^2/(L_1^2 + |\bp_1-\bp_0|^2)^2.
	\label{eq:rs-sta}
\end{equation}
This accords perfectly with our intuition: each point on the diffuser contributes incoherently to the final irradiance---these contributions are delayed according to the distance they travel, scaled by the inverse square of said distance, and scaled by the cosine-squared obliquity factor~\cite{footnote-hf}.

\section{The 6D light field and TFSWD}
\label{sec:tfswd}
Next we turn our attention to the TFSWD.  Here it turns out to be convenient to first seek the 6D light field's primitive for nonparaxial propagation starting from its paraxial-propagation predecessor,
\begin{align}
	I_1(\bpp,\bs,\wpp,t) &\equiv \frac{1}{\lo^2}\intw{\wm} W_{\scE_1}(\bpp,2\pi\bs/\lo,\wpp,\wm) e^{-i\wm t} \\
	&= \frac{1}{\lo^2}\intw{\wm} W_{\scE_{0}'}(\bpp - L_1\bs,k_0 \bs,\wpp,\wm) e^{i\wm L_1\parens{1+\abs{\bs}^2/2}/c}e^{-i\wm t} \\
	&= I_{0}'(\bpp-L_1\bs,\bs,\wpp,t-L_1/c-\abs{\bs}^2L_1/2c).
\end{align}
Interpreting the first argument in $I_0'$ as a transverse spatial coordinate $\bp_0$ in the $z=0$ plane, we see that $\bs=\parens{\bpp-\bp_0}/L_1$. This identification makes sense in the paraxial regime: $\bs$ represents the transverse component of the propagation direction, and so it equals the ratio of the transverse spatial offset to the paraxial-propagation distance. Hence, the 6D light field's spatial profile is merely sheared according to the propagation direction. The propagation direction itself does not change, nor does the frequency, and we see that the time dependence is delayed by light-speed propagation over the paraxial-propagation distance $L_1 + \abs{\bs}^2 L_1/2 = L_1+\abs{\bpp-\bp_0}^2/2L_1$. This interpretation demonstrates that the 6D light field formalizes the ray-optics intuition for free-space propagation.

\begin{figure}[hbt]
	\centering
	\includegraphics[width=3in]{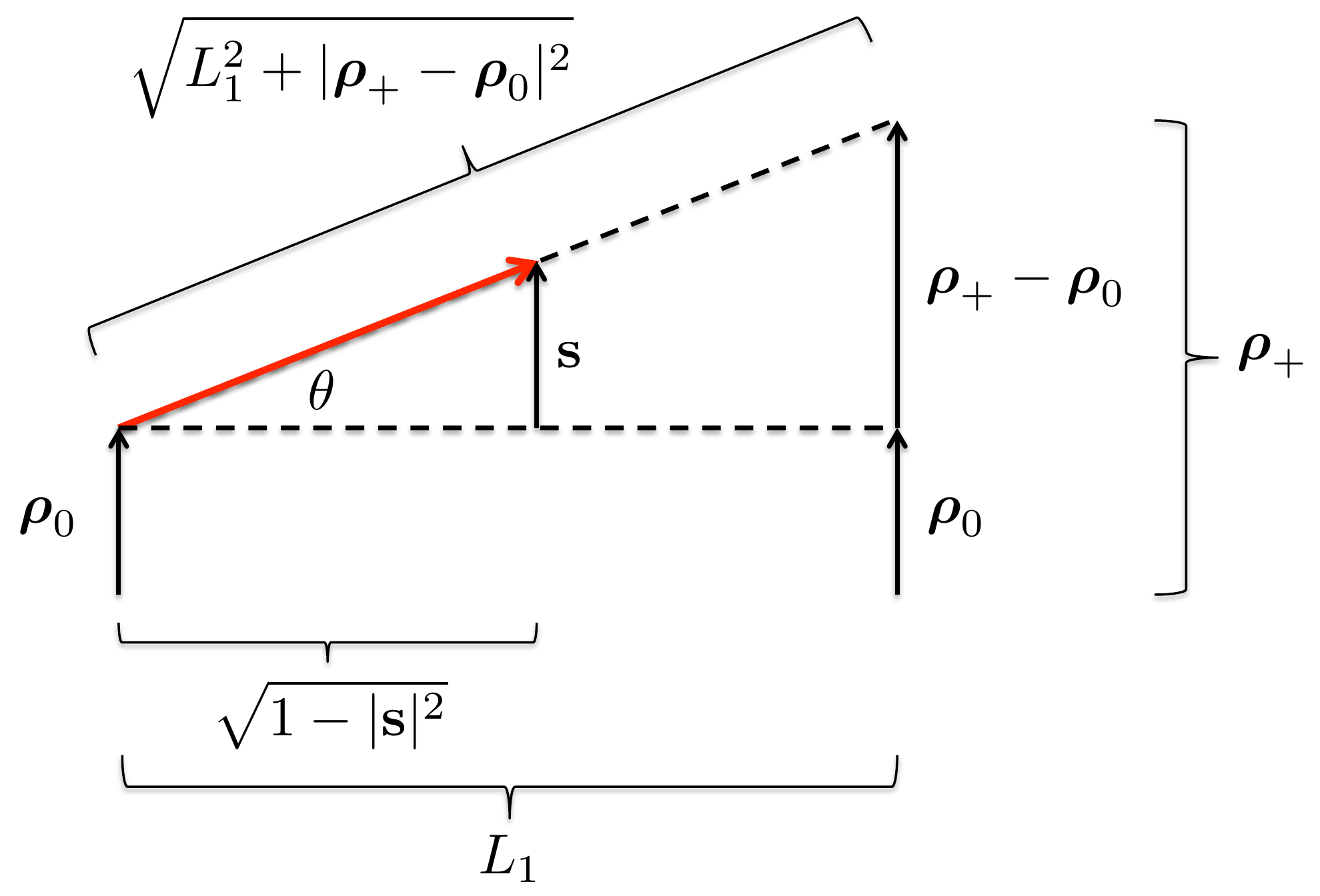}
	\caption{Trigonometry for Rayleigh--Sommerfeld propagation of the 6D light field. The red arrow represents a unit vector pointing in the propagation direction of a hypothetical ray propagating from $(\bp_0,0)$ to $(\bp_+,L_1)$.
		\label{fig:ray-optics}}
\end{figure}

The preceding interpretation immediately exposes the imprecision of the Fresnel approximation. Neither $L_1$, as in the denominator of our interpretation for $\bs$, nor $L_1+\abs{\bpp-\bp_0}^2/2L_1$, as in the time delay, are really the propagation distance of such a hypothetical ray. That distance would be $\sqrt{L_1^2+\abs{\bpp-\bp_0}^2}$. Say then we take $\bs = \parens{\bpp-\bp_0}/\sqrt{L_1^2+\abs{\bpp-\bp_0}^2}$ and solve for $\bp_0$. We find that
\begin{align}
	\bp_0 &= \bpp - \frac{L_1\bs}{\sqrt{1-\abs{\bs}^2}}.
	\label{eq:ray-last}
\end{align}
Equation~(\ref{eq:ray-last}) accords well with intuition. Define $\theta$ to be the angle the propagation direction makes with the $z$ axis, as depicted in Fig.~\ref{fig:ray-optics}. Trigonometrically, $L_1$ represents the adjacent arm's length in that figure's dashed triangle.  Because the propagation direction is a unit vector, and $\bs$ is its transverse component, we have $\abs{\bs}=\sin\theta$, $\sqrt{1-\abs{\bs}^2}=\cos\theta$, and  $\abs{\bs}/\sqrt{1-\abs{\bs}^2} = \tan\theta$. We then have that $L_1\abs{\bs}/\sqrt{1-\abs{\bs}^2}$ is the transverse propagation distance. Removing the magnitude bars from the numerator gives the vector-valued transverse displacement in the correct direction, and the total propagation distance is then $\sqrt{L_1^2+\abs{\bpp-\bp_0}^2} = L_1/\sqrt{1-\abs{\bs}^2}$.

Taking the appropriately corrected values for the transverse offset and the total time delay, this suggests that the correct free-space propagation primitive for the 6D light field is
\begin{align}
	I_1(\bpp,\bs,\wpp,t) = I'_{0}\parens{\bpp-\frac{L_1\bs}{\sqrt{1-\abs{\bs}^2}},\bs,\wpp,t-\frac{L_1}{c\sqrt{1-\abs{\bs}^2}}}.
	\label{eq:6D-prim}
\end{align}
To check this result, we can ask what it implies about the propagation of the diffuser-averaged STA irradiance following a diffuser. It follows from the diffuser TFSWD primitive, i.e., Eq.~(\ref{diffuserprim}), that
\begin{align}
	I'_{0}(\bpp,\bs,\wpp,t)=\intII{\bs'} I_0(\bpp,\bs',\wpp,t).
\end{align}
So, we have that
\begin{align}
	\avg{I_1(\bp_1,t)} &= \intw{\wpp}\intII{\bs} I_1(\bp_1,\bs,\wpp,t)\\
	&= \intw{\wpp}\intII{\bs}\intII{\bs'} I_0\parens{\bp_1-\frac{L_1\bs}{\sqrt{1-\abs{\bs}^2}},\bs',\wpp,t-\frac{L_1}{c\sqrt{1-\abs{\bs}^2}}}.
\end{align}
Now we change variables so that $\bs = \parens{\bp_1-\bp_0}/\sqrt{L_1^2+\abs{\bp_1-\bp_0}^2}$. Computing the appropriate Jacobian determinant we find $\intII{\bs}=\intII{\bp_0} L_1^2/\parens{L_1^2+\abs{\bp_1-\bp_0}^2}^2$, so we now have
\begin{align}
	\avg{I_1(\bp_1,t)} =& \intw{\wpp}\intII{\bp_0}\frac{L_1^2}{\parens{L_1^2+\abs{\bp_1-\bp_0}^2}^2}\nonumber\\&\times\intII{\bs'}I_0\parens{\bp_0,\bs',\wpp,t-\frac{\sqrt{L_1^2+\abs{\bp_1-\bp_0}^2}}{c}}\\
	=&\intII{\bp_0}I_0\parens{\bp_0,t-\frac{\sqrt{L_1^2+\abs{\bp_1-\bp_0}^2}}{c}}\frac{L_1^2}{\parens{L_1^2+\abs{\bp_1-\bp_0}^2}^2},
\end{align}
which is precisely our result for Rayleigh--Sommerfeld propagation of the diffuser-averaged STA irradiance after a diffuser, Eq.~(\ref{eq:rs-sta}). Moreover, note how the cosine-squared obliquity factor is automatically accounted for by this approach.

Inspired by this success, we propose the equivalent nonparaxial free-space propagation primitive for the TFSWD,
\begin{align}
	W_{\scE_1}(\bpp,\bk,\wpp,\wm) = W_{\scE_{0}'}\parens{\bpp-\frac{L_1\bk}{\sqrt{k_0^2-\abs{\bk}^2}},\bk,\wpp,\wm}e^{i\wm L_1 k_0 / c\sqrt{k_0^2-\abs{\bk}^2}},
	\label{eq:tfswd-prim}
\end{align}
which can be obtained by Fourier transforming Eq.~(\ref{eq:6D-prim}). In principle, one would hope to derive this formula directly from substituting the Rayleigh--Sommerfeld diffraction integral for the complex field envelope into the definition of the TFSWD. Unfortunately, we find this approach untenable. However, this result can be derived under special circumstances by adapting a technique developed elsewhere for the traditional optical Wigner distribution~\cite{Alonso2011}, as we will now show for free-space propagation from the 0 plane to the $z$ plane.    

We begin that demonstration by shifting our attention from the output plane's complex field envelope $\scE_z(\bp_z,\omega)$ to its angular spectrum $\scEt_z(\bs,\omega)$, in which the field envelope is expressed as a superposition of propagating plane waves and all evanescent components are ignored, i.e.,
\begin{align}
	\scE_z(\bp_z,\omega) &= \int\!\frac{{\rm d}^2\bs}{(2\pi)^2} \scEt_z(\bs,\omega) \exp[i\parens{\wo+\omega}\bp_z\cdot\bs/c]\\
	\scEt_z(\bs,\omega) &= \scEt'_0(\bs,\omega) \exp[i\parens{\wo+\omega}z s_z(\bs)/c],
\end{align}
where $|\scEt_z(\bs,\omega)| = 0$ for $|\bs| > 1$, $s_z(\bs)\equiv\sqrt{1-\abs{\bs}^2}$ is real valued when the integrand is nonzero, and the integration limits can be taken to include the entire $\bs$ plane.  Substituting these results into the TFSWD's definition we have that
\begin{align}
	&W_{\scE_z}(\bpp,\bk,\wpp,\wm)\nonumber\\ &= \intII{\bpm} e^{-i\bk\cdot\bpm} \avg{\scE_z(\bpp+\bpm/2,\wpp+\wm/2) \scE_z^*(\bpp-\bpm/2,\wpp-\wm/2)} \\
	&= \intII{\bpm} \intk{\bs} \intk{\bst} e^{-i\bk\cdot\bpm} e^{i\parens{\wo+\wpp+\wm/2}[(\bpp+\bpm/2)\cdot\bs+z s_z(\bs)]/c} \nonumber\\&\ \times e^{-i\parens{\wo+\wpp-\wm/2}[(\bpp-\bpm/2)\cdot\bst+z s_z(\bst)]/c}  \avg{\scEt'_0(\bs,\wpp+\wm/2)\scEt_0^{\prime *}(\bst,\wpp-\wm/2)}.
\end{align}
Now we switch to sum and difference coordinates, $\bsp=\parens{\bs+\bst}/2$ and $\bsm = \bs - \bst$. After reorganizing some terms we have that
\begin{align}
	W_{\scE_z}(\bpp,\bk,\wpp,\wm) =& \intII\bpm \intk\bsp \intk\bsm \nonumber\\&\times e^{i\bpm\cdot[\parens{\wo+\wpp}\bsp/c+\wm\bsm/4c - \bk]} e^{i\bpp\cdot[\parens{\wo+\wpp}\bsm/c+\wm\bsp/c]} \nonumber\\ &\times e^{i\parens{\wo+\wpp}z\bracket{s_z(\bsp+\bsm/2)-s_z(\bsp-\bsm/2)}/c} e^{i\wm z\bracket{s_z(\bsp+\bsm/2)+s_z(\bsp-\bsm/2)}/2c} \nonumber\\&\times\avg{\scEt'_0(\bsp+\bsm/2,\wpp+\wm/2)\scEt_0^{\prime *}(\bsp-\bsm/2,\wpp-\wm/2)}.
\end{align}
Performing the $\bpm$ integral yields
\begin{align}
	\intII\bpm e^{i\bpm\cdot[\parens{\wo+\wpp}\bsp/c+\wm\bsm/4c - \bk]} = \parens{\frac{2\pi c}{\wo+\wpp}}^2\delta\!\parens{\bsp-\frac{c\bk}{\wo+\wpp}+\frac{\wm \bsm}{4\parens{\wo+\wpp}}},
\end{align}
where the last term in the argument of the delta function can be ignored owing to $|\bs_-| \le 2$ for propagating waves and our quasimonochromatic assumption, viz., that $\scE_z(\bp_z,\omega)\sim 0$ except for $|\omega| \ll \wo$ and hence $|\wm|,|\wpp|\ll\wo$. Defining $\kot \equiv (\wo+\wpp)/c$ and evaluating the $\bsp$ integral then yields
\begin{align}
	W_{\scE_z}(\bpp,\bk,\wpp,\wm) =& \intk\bsm  \frac{1}{\kot^2}e^{i\bpp\cdot\parens{\kot\bsm+\wm\bk/c\kot}}  e^{i\kot z[s_z(\bk/\kot+\bsm/2)-s_z(\bk/\kot-\bsm/2)]} \nonumber\\&\times e^{i\wm z[s_z(\bk/\kot+\bsm/2)+s_z(\bk/\kot-\bsm/2)]/2c} \nonumber\\&\times\avg{\scEt'_0(\bk/\kot+\bsm/2,\wpp+\wm/2)\scEt_0^{\prime *}(\bk/\kot-\bsm/2,\wpp-\wm/2)}.
	\label{exact}
\end{align}

Next we introduce the first of our two key assumptions. Consider the case in which the angular-spectrum correlation $\avg{\scEt'_0(\bsp+\bsm/2,\wpp+\wm/2)\scEt_0^{\prime *}(\bsp-\bsm/2,\wpp-\wm/2)}$ differs appreciably from 0 only when $\abs{\bsm} \ll 1$. In that case, we can approximate $s_z$ by its first-order expansion in $\bsm$, i.e., $s_z(\bsp\pm\bsm/2)\approx s_z(\bsp) \pm \nabla s_z(\bsp)\cdot \bsm/2$ where $\nabla s_z(\bsp) = -\bsp/s_z(\bsp)$. Using this approximation and reorganizing terms we now have
\begin{align}
	W_{\scE_z}(\bpp,\bk,\wpp,\wm) =& \intk\bsm \frac{1}{\kot^2} e^{i\kot\bsm\cdot[z \nabla s_z(\bk/\kot) + \bpp]} e^{i\wm[z s_z(\bk/\kot)+\bpp\cdot\bk/\kot]/c}\nonumber\\&\times \avg{\scEt'_0(\bk/\kot+\bsm/2,\wpp+\wm/2)\scEt_0^{\prime *}(\bk/\kot-\bsm/2,\wpp-\wm/2)}.
\end{align}
Because the angular spectrum can be obtained from the complex field envelope via
\begin{align}
	\scEt_0'(\bs,\omega) = \intII{\bp_0} \kot^2\scE_0'(\bp_0,\omega) \exp[-i\parens{\wo+\omega}\bp_0\cdot\bs/c],
	\label{assump1}
\end{align}
where $\scE_0'(\bp_0,\omega)$ only includes propagating waves, we find that
\begin{align}
	W_{\scE_z}(\bpp,\bk,\wpp,\wm) =& \intII{\bp_0} \intII{\bpt_0} \intk\bsm \kot^2 \nonumber\\&\times e^{i\kot\bsm\cdot[z \nabla s_z(\bk/\kot) + \bpp]} e^{i\wm[z s_z(\bk/\kot)+\bpp\cdot\bk/\kot]/c}\nonumber\\&\times e^{-i\parens{\wo+\wpp+\wm/2}\bp_0\cdot\parens{\bk/\kot + \bsm/2}/c}e^{i\parens{\wo+\wpp-\wm/2}\bpt_0\cdot\parens{\bk/\kot - \bsm/2}/c}\nonumber\\&\times \avg{\scE'_0(\bp_0,\wpp+\wm/2)\scE_0^{\prime *}(\bpt_0,\wpp-\wm/2)}.
\end{align}
Changing to sum and difference coordinates again and reorganizing terms we have
\begin{align}
	W_{\scE_z}(\bpp,\bk,\wpp,\wm) =& \intII{\bpt_+} \intII{\bpt_-} \intk\bsm \kot^2 e^{-i\kot\bsm\cdot\{\bpt_+-[\bpp+z \nabla s_z(\bk/\kot)]+\wm\bpt_-/4(\wo+\wpp)\}} \nonumber\\&\times e^{-i\bk\cdot\bpt_-} e^{i\wm\parens{z s_z(\bk/\kot) +(\bpp-\bpt_+)\cdot\bk/\kot}/c}\nonumber\\&\times \avg{\scE'_0(\bpt_++\bpt_-/2,\wpp+\wm/2)\scE_0^{\prime *}(\bpt_+-\bpt_-/2,\wpp-\wm/2)}.
\end{align}
Now, because the integrand only includes propagating waves, the $\bsm$ integral can be done over the entire ${\bf s}_-$ plane, yielding
\begin{align}
	\intk\bsm \kot^2 &e^{-i\kot\bsm\cdot\{\bpt_+-[\bpp+z \nabla s_z(\bk/\kot)]+\wm\bpt_- / 4(\wo+\wpp)\}}\nonumber\\ &= \delta \parens{\bpt_+-[\bpp+z \nabla s_z(\bk/\kot)]+\frac{\wm\bpt_-}{4(\wo+\wpp)}}.
\end{align}

Here we introduce the second of our key assumptions, that the complex-field-envelope correlation $\avg{\scE'_0(\bpt_++\bpt_-/2,\wpp+\wm/2)\scE_0^{\prime *}(\bpt_+-\bpt_-/2,\wpp-\wm/2)}$ differs significantly from 0 only for $\abs{\bpt_-}$ small enough that our quasimonochromatic assumption suffices for us to neglect the final term in the argument of the preceding delta function. In this case, we carry out the $\bpt_+$ integral with the help of the delta function and get
\begin{align}
	W_{\scE_z}(\bpp,\bk,\wpp,\wm) =& \intII{\bpt_-} e^{-i\bk\cdot\bpt_-} e^{i\wm z [s_z(\bk/\kot) - \nabla s_z(\bk/\kot)\cdot\bk/\kot]/c}\nonumber\\&\times \langle\scE'_0(\bpp+z \nabla s_z(\bk/\kot)+\bpt_-/2,\wpp+\wm/2)\nonumber\\&\times\scE_0^{\prime *}(\bpp+z \nabla s_z(\bk/\kot)-\bpt_-/2,\wpp-\wm/2)\rangle 
\end{align}
\begin{align}
	\hspace{0.5in}
	=&\,\, W_{\scE'_0}(\bpp+z \nabla s_z(\bk/\kot),\bk,\wpp,\wm) e^{i \wm z [s_z(\bk/\kot)-\nabla s_z(\bk/\kot)\cdot\bk/\kot]/c} \\
	=&\,\, W_{\scE'_0}\parens{\bpp-\frac{z\bk}{\sqrt{\kot^2-\abs{\bk}^2}},\bk,\wpp,\wm} e^{i \wm z \kot / c \sqrt{\kot^2-\abs{\bk}^2}}.
\end{align}
Finally, we make use of our quasimonochromatic assumption one more time to say $\kot\approx\ko$ which gives us the final result we desire:
\begin{align}
	W_{\scE_z}(\bpp,\bk,\wpp,\wm) = W_{\scE'_0}\parens{\bpp-\frac{z\bk}{\sqrt{\ko^2-\abs{\bk}^2}},\bk,\wpp,\wm} e^{i\wm z \ko / c \sqrt{\ko^2-\abs{\bk}^2}}.
	\label{eq:rs-wigner}
\end{align}

Before moving on, let us restate and more carefully consider our two key assumptions: (1) that $\avg{\scEt'_0(\bs,\omega)\scEt_0^{\prime *}(\bst,\tilde\omega)}$ differs appreciably from 0 only when $\abs{\bs-\bst} \ll 1$; and (2) that $\avg{\scE'_0(\bp_0,\omega)\scE_0^{\prime *}(\bpt_0,\tilde\omega)}$ differs significantly from 0 only for $\abs{\bp_0-\bpt_0}$ small enough that $\abs{(\bp_0+\bpt_0)/2-z\bk/\sqrt{\ko^2-\abs{\bk}^2}} \gg (\omega-\tilde{\omega}) \abs{\bp_0-\bpt_0} / 4[\wo+(\omega + \tilde{\omega})/2]$. We will make that consideration when $E_0(\bp_0,t)$ is space-time factorable, i.e., when $E_0(\bp_0,t) = E_0(\bp_0)S(t)$, so that the complex field envelope emerging from the 0 plane and its temporal Fourier transform are
\begin{align}
	E'_0(\bp_0,t) = E'_0(\bp_0) S(t)\\
 	\scE'_0(\bp_0,\omega) = E'_0(\bp_0) \scS(\omega),
\end{align}
where $\scS(\omega) \equiv \int\!{\rm d}t\,S(t)e^{i\omega t}$.  If the 0 plane in this discussion is Fig.~\ref{fig:geo}'s $z=0$ plane we know that $E_0'(\bp_0) \approx e^{i\omega_0h_0(\bp_0)/c}E_0(\bp_0)$, where $E_0(\bp_0)$ is deterministic.  Equation~(\ref{eq:diff-stat}) then immediately validates our second key assumption and we also have that
\begin{equation}
\langle \scEt'_0(\bs,\omega)\scEt_0^{\prime *}(\bst,\tilde\omega)\rangle = \scS(\omega)\scS^*(\tilde{\omega})\int\!{\rm d}^2\bp_0\,\lo^2\tilde{k}_0^4I_0(\bp_0)e^{-i\omega_0\bp_0\cdot[(\bs-\bst)+\hat{\bs}]/c},
\end{equation}
where $\hat{\bs}\equiv (\bs\omega-\bst\tilde{\omega})/\omega_0$ satisfies $|\hat{\bs}|\ll 1$ because $|\omega|,|\tilde{\omega}|\ll \omega_0$ and $|\bs|,|\bst|<1$. Thus our first assumption will be satisfied if the angular spectrum of $I_0(\bp_0)$ evaluated at $\bs-\bst + \hat{\bs}$ is confined to the region $|\bs-\bst+\hat{\bs}|\ll 1$.  This condition \emph{does not} require the field itself to be paraxial, i.e., it does not presume $|\bs|,|\bst|\ll 1$.   

Unfortunately, the preceding argument does not suffice for validating Eq.~(\ref{eq:tfswd-prim}) as the nonparaxial replacement for Eq.~(\ref{Fresnelprim}) because the latter applies to \emph{all} the free-space propagation paths in Fig.~\ref{fig:geo}.  In particular, Eq.~(\ref{Fresnelprim})'s derivation in Ref.~\cite{Dove2019} does \emph{not} require the input TFSWD to be that for a field emerging from a diffuser.  Hence our paraxial-propagation primitive for the TFSWD includes free-space propagation from the $z=L_1$ and $z=L_1+L_d$ planes in Fig.~\ref{fig:geo}, whereas what we have proven so far for its nonparaxial replacement does not.  To show that Eq.~(\ref{eq:tfswd-prim}) can be applied to those non-diffuser situations we must work harder by allowing $E_0'(\bp_0)$'s correlation function to be more general than the $\delta$-function form implied by Eq.~(\ref{eq:diff-stat}).  In that case---using $\bp_+ = (\bp_0+\bpt_0)/2$, $\bp_- = \bp_0-\bpt_0$, $\bs_+ = (\bs + \bst)/2$, $\bs_-= \bs-\bst$, $\omega = \wpp+\wm/2$ and $\tilde\omega = \wpp - \wm/2$---we have that
\begin{align}
	\avg{\scEt'_0(\bs,\omega)\scEt_0^{\prime *}(\bst,\tilde\omega)} =& \scS(\omega)\scS^*(\tilde\omega)\intII{\bpp}\intII{\bpm} \avg{E'_0(\bp_0)E^{\prime *}_0(\bpt_0)} \nonumber\\&\times e^{-i\bpp\cdot[\wm\bsp+(\wo+\wpp)\bsm]/c} e^{-i\bpm\cdot[\wm\bsm/4+(\wo+\wpp)\bsp]/c},
\end{align}
and
\begin{align}
	\avg{\scE'_0(\bp_0,\omega)\scE_0^{\prime *}(\bpt_0,\tilde\omega)} = \scS(\omega)\scS^*(\tilde\omega)\avg{E'_0(\bp_0)E^{\prime *}_0(\bpt_0)}.
\end{align}
Next, let us assume that the spatial correlation function has a Schell model, viz.,
\begin{align}
	\avg{E'_0(\bp_0)E'^*_0(\bpt_0)} = \langle I'_0(\bpp)\rangle R(\bpm),
\end{align}
where, as its notation suggests, $\langle I'_0(\bpp)\rangle \equiv \langle |E'_0(\bpp)|^2\rangle$ and $R(\bpm)$ is a normalized ($R({\bf 0}) = 1$), spatially-homogeneous correlation function. In this case,
\begin{align}
	\avg{\scEt'_0(\bs,\omega)\scEt_0^{\prime *}(\bst,\tilde\omega)} =&\,\, \scS(\omega)\scS^*(\tilde\omega)\intII{\bpp}\intII{\bpm} \langle I'_0(\bpp)\rangle R(\bpm) \nonumber\\&\times e^{-i\bpp\cdot[\wm\bsp+(\wo+\wpp)\bsm]/c} e^{-i\bpm\cdot[\wm\bsm/4+(\wo+\wpp)\bsp]/c}\\
	=&\,\, \scS(\omega)\scS^*(\tilde\omega) \scI\!\parens{\bracket{\wm\bsp+(\wo+\wpp)\bsm}/c}\nonumber\\&\times \scR\!\parens{\bracket{\wm\bsm/4+(\wo+\wpp)\bsp}/c}\\
	\approx&\,\, \scS(\omega)\scS^*(\tilde\omega) \scI(\ko\bsm) \scR(\ko\bsp),
\end{align}
where 
\begin{align}
	\scI(\bk) &\equiv \intII{\bp} \langle I'_0(\bp)\rangle \exp(-i\bp\cdot\bk)\\
	\scR(\bk) &\equiv \intII{\bp} R(\bp) \exp(-i\bp\cdot\bk),
\end{align}
and the approximation follows from $|\bsm|/4 \le 1/2$, $|\bsp|\le1$, and $|\wm|, |\wpp| \ll \wo$.
So, our two assumptions then amount to: (1) $\langle I'_0(\bpp)\rangle$ having its $\scI(\bk)$ confined to a sufficiently narrow region about $\bk = {\bf 0}$; and (2) $R(\bpm)$ being confined to a sufficiently narrow region about $\bpm = {\bf 0}$. These conditions generalize what we stated earlier when we noted that the 0 plane's containing a diffuser as in Fig.~\ref{fig:geo} leads to our two key assumptions being easily satisfied.  

To gain further insight into the preceding partial-coherence validity conditions for Eq.~(\ref{Fresnelprim}), suppose that 
\begin{align}
	\langle I'_0(\bpp)\rangle &= \langle I'_0(\bmu)\rangle\exp(-\abs{\bpp-\bmu}^2/2\rho_I^2) \\
	R(\bpm) &= \exp(-\abs{\bpm}^2/2\rho_R^2),
\end{align}
where $\rho_I$ and $\rho_R$ are the functions' $e^{-1/2}$-attenuation radii. These functions lead to
\begin{align}
	\scI(\bk) &= 2\pi\rho_I^2 \langle I'_0(\bmu)\rangle\exp(-\rho_I^2\abs{\bk}^2/2-i\bk\cdot\bmu)\\
	\scR(\bk) &= 2\pi\rho_R^2  \exp(-\rho_R^2\abs{\bk}^2/2).
\end{align}
From these expressions it follows that our assumptions will be satisfied if (1) $\rho_I \gg \lambda_0$ and (2) $\abs{\bpp-z\bk/\sqrt{\ko^2-\abs{\bk}^2}} \gg \wm \rho_R / \wo$. 
The first constraint will be easily satisfied for any 0-plane configuration of interest. Understanding the second constraint requires more thought. Recalling our geometric interpretation from Fig.~\ref{fig:ray-optics},  Eq.~(\ref{eq:ray-last}) implies that a $\bk$-propagating ray that crosses the 0 plane at $\bp_0$ will cross the $z$ plane at $\bpp$.  Thus, the second constraint reduces to $|\bp_0| \gg \wm \rho_R / \wo$, and hence Eq.~({\ref{eq:rs-wigner}}) is valid except for $(\bpp$,$\bk)$ pairs that originate within a small radius about the 0 plane's origin. For the case of a partial diffuser, this radius is given by a small fraction of its coherence length. Taking $\wm/2\pi = 10\,\text{GHz}$ and $\lambda_0=532\nm$, we have $\wm/\wo\approx 1.8\times10^{-5}$. So, our results should be valid for any rays that originate outside $|\bp_0|\gg \rho_R / 56000$.

To make the preceding analysis applicable to interaction with an occluder, suppose that the frequency-domain envelope $\scE_0'(\bp_0,\omega)$ illuminates an occluder $P(\bp_0)$ in the 0 plane resulting in a frequency-domain envelope $\scE_0''(\bp_0,\omega) = \scE_0'(\bp_0,\omega)P(\bp_0)$, where, e.g., $P(\bp_0)$ could be the Gaussian pinspeck considered in Ref.~\cite{Dove2019}'s analysis of occlusion-aided $\scP$-field imaging.  This new frequency-domain envelope's correlation function is then
\begin{equation}
\avg{\scE_0''(\bp_0,\omega)\scE_0''(\bpt_0,\tilde{\omega})} = \scS(\omega)\scS^*(\tilde\omega)\langle I'_0(\bpp)\rangle R(\bpm)P(\bp_0)P^*(\bpt_0).
\end{equation}
If $P(\bp_0)P^*(\bpt_0)$ varies insignificantly for $|\bp_0-\bpt_0|$ less than the correlation width of $R(\bpm)$, we get  
\begin{equation}
\avg{\scE_0''(\bp_0,\omega)\scE_0''(\bpt_0,\tilde{\omega})} \approx \scS(\omega)\scS^*(\tilde\omega)\langle I'_0(\bpp)\rangle R(\bpm)|P(\bpp)|^2,
\end{equation}
and the validity conditions identified in the previous paragraph apply with $I''_0(\bpp) \equiv I'_0(\bpp)|P(\bpp)|^2$ used in place of $I'_0(\bpp)$.  

\section{The Helmholtz equation}
\label{sec:helmholtz}

The last section required considerable work to obtain and validate---under some restrictions---a nonparaxial free-space propagation primitive for the TFSWD. So, it is worth seeking an alternative, differential-equation characterization that might provide a better route to such a primitive. In terms of the underlying positive-frequency optical field---which, in the time domain we denote $U_z^{(+)}(\bp_z,t) \equiv E_z(\bp_z,t) e^{-i\wo t}$---the most general description we can provide is the wave equation,
\begin{align}
	\parens{\nabla_{\bp_z}^2+\partial_z^2-\frac{1}{c^2}\partial_t^2}U_z^{(+)}(\bp_z,t) = 0,
\end{align}
where $\nabla_{\bp_z}$ is the 2D gradient with respect to the transverse coordinate $\bp_z$ and $\nabla_{\bp_z}^2$ is the associated Laplacian. The implication for the frequency-domain complex field envelope is that it satisfies the Helmholtz equation with a frequency offset,
\begin{align}
	\parens{\nabla_{\bp_z}^2+\partial_z^2-\frac{(\wo+\omega)^2}{c^2}}\scE_z(\bp_z,\omega) = 0.
	\label{eq:scEhelm}
\end{align}
One might expect that the Rayleigh--Sommerfeld diffraction integral, Eq.~(\ref{eq:RS1}), would satisfy this equation with $\scE'_0(\bp_0,\omega)$ as a boundary condition. However, it does not, as the Rayleigh--Sommerfeld integral is in fact an approximation of the complete diffraction integral we will call the Rayleigh diffraction integral~\cite{mandw},
\begin{align}
	\scE_z(\bp_z,\omega) =& \intII{\bp_0} \scE'_0(\bp_0,\omega) \frac{1}{2\pi}\parens{\frac{1}{\sqrt{z^2+\abs{\bp_z-\bp_0}^2}}-i\frac{\wo+\omega}{c}}\nonumber\\&\times\frac{z}{z^2+\abs{\bp_z-\bp_0}^2} e^{i\parens{\wo+\omega}\sqrt{z^2+\abs{\bp_z-\bp_0}^2}/c},
	\label{eq:rayleigh-diff}
\end{align}
which, as it turns out, \emph{does} satisfy Eq.~(\ref{eq:scEhelm}) with $\scE'_0(\bp_0,\omega)$ as a boundary condition. Equation~(\ref{eq:RS1}) is obtained from Eq.~(\ref{eq:rayleigh-diff}) in the limit of $\lambda_0\ll z$, which is trivially true for any scenario of practical interest. Though more complete than the Rayleigh--Sommerfeld diffraction integral, the Rayleigh diffraction integral is even more cumbersome to manage, and it does not lend itself as readily to intuitive physical interpretation. Accordingly, we are at a loss to find an appropriate TFSWD primitive from Rayleigh diffraction. Nevertheless, it is not hard to see that the implied $\scP$-field post-diffuser propagation is given by
\begin{align}
	\scP_1(\bp_1,\wm) =& \intII{\bp_0} \scP_0(\bp_0,\wm) \parens{1+\frac{\lambda_0^2}{4\pi^2\parens{L_1^2+\abs{\bp_1-\bp_0}^2}}}\nonumber\\&\times\frac{L_1^2}{\parens{L_1^2+\abs{\bp_1-\bp_0}^2}^2} e^{i\wm\sqrt{L_1^2+\abs{\bp_1-\bp_0}^2}/c},
\end{align}
which leads to the following result for the diffuser-averaged STA irradiance,
\begin{align}
	\avg{I_1(\bp_1,t)} =& \intII{\bp_0} I_0\parens{\bp_0,t-\sqrt{L_1^2+\abs{\bp_1-\bp_0}^2}/c} \parens{1+\frac{\lambda_0^2}{4\pi^2\parens{L_1^2+\abs{\bp_1-\bp_0}^2}}}\nonumber\\&\times\frac{L_1^2}{\parens{L_1^2+\abs{\bp_1-\bp_0}^2}^2}.
\end{align}

Despite its not being clear what the unrestricted TFSWD free-space propagation primitive would be, we \emph{can} characterize the TFSWD's behavior quite generally. First we note that the TFSWD is a Fourier transform of the complex-field-envelope correlation $\avg{\scE_z(\bp_z,\omega)\scE_z^*(\bpt_z,\tilde\omega)}$. It is not hard to see that this correlation must obey two Helmholtz equations, one for the first transverse spatial coordinate and frequency, and another for the second pair of coordinates. From the chain rule and the Fourier transform's linearity, we can easily derive a pair of differential equations for the TFSWD which can be summarized by
\begin{align}
	\parens{\frac{1}{4}\nabla_{\bpp}^2+\partial_z^2\pm i \bk\cdot\nabla_{\bpp} + \frac{\parens{\wo+\wpp\pm\wm/2}^2}{c^2}-\abs{\bk}^2}W_{\scE_z}(\bpp,\bk,\wpp,\wm)=0.
\end{align}
If we define an angular spectrum for the TFSWD by Fourier transforming its transverse spatial coordinate, viz.,
\begin{align}
	\tilde{W}_{\scE_z}(\bk_+,\bk_-,\wpp,\wm) \equiv \intII{\bpp} W_{\scE_z}(\bpp,\bk_+,\wpp,\wm) e^{-i\bk_-\cdot\bpp},
\end{align}
then the equivalent pair of equations for this new quantity are
\begin{align}
	\parens{\partial_z^2+\frac{\parens{\wo+\wpp\pm\wm/2}^2}{c^2}-\abs{\bk_+\pm\bk_-/2}^2}\tilde{W}_{\scE_z}(\bk_+,\bk_-,\wpp,\wm)=0.
\end{align}

Not surprisingly, Eq.~(\ref{eq:tfswd-prim}) does not satisfy these differential equations. After all, it only suffices to reproduce the approximate Rayleigh--Sommerfeld behavior of the $\scP$ field and not the full Rayleigh-diffraction behavior. However, it seems unlikely that any clean propagation primitive will satisfy these equations exactly, as all of our propagation derivations thus far have made liberal use of the quasimonochromatic assumption, which is of course not captured by the Helmholtz equation. Still, it is intellectually satisfying to have a general description in terms of such differential equations, and they may prove useful in some presently unforeseen circumstances.

\section{Summary and discussion}
\label{sec:summary}

Motivated by initial proposals and experimental demonstrations that established the utility of the $\scP$-field concept for NLoS imaging, we embarked on a program to flesh out its theoretical understanding, beginning with study of paraxial, transmissive geometries. In this paper, mindful of the limitations of our earlier work's paraxial assumption, we have derived nonparaxial post-diffuser propagation primitives for the $\scP$ field and, equivalently, the diffuser-averaged STA irradiance using the Rayleigh--Sommerfeld diffraction integral for the complex field envelope. We have also proposed nonparaxial free-space propagation primitives for the 6D light field and, equivalently, the TFSWD, which enable handling a wider variety of NLoS scenarios than is possible using the $\scP$ field alone. Our proposal was inspired by applying geometric intuition to our result for the Fresnel limit, and its validity is suggested by the fact that it reproduces our derived propagation result for the diffuser-averaged STA irradiance. We then provided a more formal derivation of this primitive using the angular-spectrum representation of the complex field envelope under certain reasonable constraints on the relevant correlation functions. An example in which our TFSWD propagation primitive will fail is as follows.  Consider the spatial-Wigner-distribution limit of our Rayleigh--Sommerfeld diffraction primitive, viz.,
\begin{equation}
W_{\scE_z}(\bpp,\bk,0,0) = W_{\scE_0'}\!\left(\bpp-\frac{z\bk}{\sqrt{k_0^2-|\bk|^2}},\bk,0,0\right).
\label{approx}
\end{equation}
The exact result for this propagation problem in the general case of nonparaxial $\scEt'_0(\bs,0)$ can be shown, from Eqs.~(\ref{exact}), (\ref{assump1}), and (\ref{TFSWDdefn}), to be
\begin{align}
W_{\scE_z}(\bpp,\bk,0,0) = &\int\!{\rm d}^2\bpt_+\,W_{\scE_0'}(\bpt_+,\bk,0,0) \intk\bsm k_0^2  e^{ik_0 z[s_z(\bk/k_0+\bsm/2)-s_z(\bk/k_0-\bsm/2)]} \nonumber\\
&\times e^{-ik_0\bsm\cdot(\bpt_+-\bpp)}.
\label{exact2}
\end{align}
Because $s_z(\bk/k_0\pm \bsm/2) = \sqrt{1-|\bk/k_0\pm \bsm/2|^2}$, Eq.~(\ref{exact2}) only reduces to the transformation shown in Eq.~(\ref{approx}) under the  first assumption we made en route to our TFSWD Rayleigh--Sommerfeld diffraction primitive, viz., that $\avg{\scEt'_0(\bk/k_0+\bsm/2,0)\scEt_0^{\prime *}(\bk/k_0-\bsm/2,0)}$ in Eq.~(\ref{exact}) differs appreciably from zero only when $|\bsm| \ll 1$.  Pushing past our derivation's assumptions, we provided a more general characterization of TFSWD behavior by presenting a pair of differential equations it obeys.

Despite this considerable progress, which successfully extends our previously developed framework to the nonparaxial regime, much work remains to be done, particularly in applying these results. In developing this extension, we have not reanalyzed the $\scP$ field's computational and physical imaging scenarios, nor have we reconsidered the effects of speckle. Our previous investigation of these topics benefited from the ability to obtain closed-form results from Gaussian integrals, a feat which will likely not be possible when using our new nonparaxial results. Consequently, it may be necessary to abandon the pursuit of closed-form insights in favor of numerical analysis and simulation for the nonparaxial case. However, in doing so, we expect to find that the essence of our results for imaging and speckle carry over nicely to the nonparaxial regime.

Even beyond these application-oriented opportunities, there is still room for further understanding of the pure theory. While our derivation of the Rayleigh--Sommerfeld propagation primitive for the TFSWD under partially-coherent circumstances is promising evidence for its validity, as is the reproduction by that primitive of the correct STA-irradiance behavior, it would be preferable to have a formal derivation of it for propagation from arbitrary planes, without restriction. A step towards this goal might be to find the Rayleigh--Sommerfeld $\scP$-field input-output relation for occluder-interrupted, post-diffuser propagation in the geometric-optics limit, as we earlier showed to be intuitive in the paraxial regime~\cite{thesis}. Finally, the pinnacle of understanding nonparaxial propagation would be to find a Rayleigh-diffraction equivalent for the TFSWD or some other propagation primitive that satisfies the Helmholtz equation or, equivalently, our new differential equations for the TFSWD. More generally, progress in this area would include any practical developments in the understanding of the TFSWD's differential equations that lead to a better, theoretically-sound understanding of the general propagation behavior of phasor-field quantities. In steering towards such broadly-applicable results, we hope that the results of this paper will serve as the foundation for transitioning our planar, transmissive framework to one that directly addresses standard, reflective NLoS geometries and the full complexities of 3D reconstruction.

\appendix
\section{6D-light-field example \label{appendix}}
Here, within the confines of our quasimonochromatic paraxial framework from Ref.~\cite{Dove2019}, we present a simple example to illustrate the interrelationships between the complex field envelope $E_0(\bp_0,t)$, the STA irradiance $I_0(\bp_0,t)$, the $\scP$ field $\scP_0(\bp_0,\wm)$, the TFSWD $W_{\scE_0}(\bp_+,\bk,\wpp,\wm)$, the 6D light field $I_0(\bp_+,\bs,\wpp,t)$, and the specific intensity $I_0(\bp_+,\bs,t)$~\cite{footnoteSpecInt}.  

For this example we assume a space-time factorable $E_0(\bp_0,t)$ comprised of a collimated Gaussian beam centered at $\bp_0 = \bmu$ with transverse propagation vector $k_0\bt$ that is multiplied by a sinusoidally-modulated Gaussian pulse in time, i.e.,
\begin{equation}
E_0(\bp_0,t) = \sqrt{I_0}\,e^{-4|\bp_0-\bmu|^2/d_0^2}e^{ik_0\bt\cdot\bp_0}e^{-t^2/T^2}\cos(\Omega t),
\end{equation}
where $0 < \Omega \ll \wo$ and $\Omega T \gg 1$.  

The STA irradiance associated with $E_0(\bp_0,t)$,
\begin{equation}
I_0(\bp_0,t) = I_0e^{-8|\bp_0-\bmu|^2/d_0^2}e^{-2t^2/T^2}\cos^2(\Omega t) = 
I_0e^{-8|\bp_0-\bmu|^2/d_0^2}e^{-2t^2/T^2}[1+\cos(2\Omega t)]/2,
\end{equation}
carries none of $E_0(\bp_0,t)$'s directionality information and its sideband frequency is double that of the field envelope.

The $\scP$ field associated with $E_0(\bp_0,t)$,
\begin{equation}
\scP_0(\bp_0,\wm) = I_0\sqrt{\frac{\pi T^2}{8}}\,e^{-8|\bp_0-\bmu|^2/d_0^2}[e^{-\wm^2T^2/8} + (e^{-(\wm+2\Omega)^2T^2/8} + e^{-(\wm-2\Omega)^2T^2/8})/2],
\end{equation}
shows the presence of double-sideband modulation at $2\Omega$ but no information about the field envelope's transverse propagation vector.  

The frequency-domain field envelope associated with $E_0(\bp_0,t)$ is
\begin{equation}
\scE_0(\bp_0,\omega) = \sqrt{I_0}\,\sqrt{\frac{\pi T^2}{4}}\,e^{-4|\bp_0-\mu|^2/d_0^2}e^{ik_0\bt\cdot\bp_0}(e^{-(\omega-\Omega)^2T^2/4}+e^{-(\omega+\Omega)^2T^2/4}).
\end{equation}

The TFSWD associated with $E_0(\bp_0,t)$, 
\begin{align}
W_{\scE_0}&(\bp_+,\bk,\wpp,\wm) = I_0\frac{\pi T^2}{4}\frac{\pi d_0^2}{2}e^{-8|\bpp-\bmu|^2/d_0^2}e^{-|\bk-k_0\bt|^2d_0^2/8} [e^{-(\wpp-\Omega)^2T^2/2}e^{-\wm^2T^2/8} \nonumber\\[.05in] 
&+ e^{-\wpp^2T^2/2}e^{-(\wm/2-\Omega)^2T^2/2}+ e^{-\wpp^2T^2/2}e^{-(\wm/2+\Omega)^2/2} + e^{-(\wpp+\Omega)^2T^2/2}e^{-\wm^2T^2/8}],
\end{align}
contains information about \emph{both} $\bt$ and $\Omega$.  

The 6D light field associated with $E_0(\bp_0,t)$,
\begin{align}
I_0(\bpp,\bs,\wpp,t) &= \frac{\pi d_0^2I_0}{8\lambda_0^2}\sqrt{2\pi T^2}\,e^{-8|\bpp-\bmu|^2/d_0^2}e^{-k_0^2|\bs-\bt|^2d_0^2/8}e^{-2t^2/T^2}[e^{-(\wpp-\Omega)^2T^2/2} \nonumber \\[.05in]
&+ 2 e^{-\wpp^2T^2/2}\cos(2\Omega t) + e^{-(\wpp+\Omega)^2T^2/2}], 
\end{align}
accords with the interpretation given in the text.  In particular, the radiance is concentrated: (1) in space around $\bp_+ = \bmu$; (2) in transverse propagation vector around $\bs = \bt$; (3) in optical frequency around $\wo+\wpp = \wo$ and $\wo+\wpp = \wo\pm \Omega$; and (4) in time around $t=0$.  

The time-dependent specific intensity associated with $E_0(\bp_0,t)$, 
\begin{align}
I_0(\bp_+,\bs,t) &= \int\!\frac{{\rm d}\wpp}{2\pi}\,I_0(\bpp,\bs,\wpp,t)\\[.05in]
&= \frac{\pi d_0^2I_0}{4\lambda_0^2}e^{-8|\bpp-\bmu|^2/d_0^2}e^{-k_0^2|\bs-\bt|^2d_0^2/8}e^{-2t^2/T^2}[1+\cos(2\Omega t)],
\end{align}
quantifies the time-$t$ radiance at $\bpp$ with transverse propagation vector $\bs$.  Integrating this result over $\bs$ then recovers the STA irradiance, i.e., we have that
\begin{equation}
\int\!{\rm d}\bs\,I_0(\bp_+,\bs,t) = 
I_0e^{-8|\bp_+-\bmu|^2/d_0^2}e^{-2t^2/T^2}[1+\cos(2\Omega t)]/2 = I_0(\bpp,t),
\end{equation}
as expected from the physical interpretations of $I_0(\bp_+,\bs,t)$ and $I_0(\bpp,t)$.
  
\section*{Funding}
Defense Advanced Research Projects Agency (HR0011-16-C-0030).

\section*{Acknowledgments}
The authors acknowledge fruitful interactions with members of the various DARPA REVEAL university teams and the REVEAL government team regarding the phasor-field concept and its applicability to NLoS imaging.

\section*{Disclosures}
The authors declare no conflicts of interest.\\


\begin{thebibliography}{99}
\bibitem{footnote-z} Throughout this paper, we will simplify subscripts involving $z$ where the meaning is clear from context.
\bibitem{footnote-speckle} When analyzing speckle effects, higher-order moments, viz., fourth-order moments in particular, will be of relevance. However, speckle analysis can often rely on Gaussian-statistics assumptions that enable calculating these moments from second-order moments via Gaussian moment factoring~\cite{speckle}.
\bibitem{footnoteSpecInt}The time-dependent specific intensity, which comes from small-angle approximation transport theory~\cite{Ishimaru}, is computer vision's 5D light field~\cite{lf1,lf2,lf3}.   It equals the radiance that is present at transverse location $\bp_+$ and time $t$ whose transverse propagation vector is $\bs$.
\end{thebibliography}
\end{document}